# Thickness and temperature-dependent damping in $La_{0.67}Sr_{0.33}MnO_3$ epitaxial films

Yifei Wang,[1] Xinxin Fan,[1] Xiaoyu Feng,[1] Xiaohu Gao,[1] Yunfei Ke,[1] Jiguang Yao,[1,2] Muhan Guo,[1] Tao Wang,[1] Lvkang Shen,[3,4] Ming Liu,[3,4] Desheng Xue,[1] and Xiaolong Fan[1]

## AFFILIATIONS

[1]Key Laboratory for Magnetism and Magnetic Materials of the Ministry of Education, Lanzhou University, Lanzhou 730000, People's Republic of China
[2]Department of Physics and Astronomy, University of Manitoba, Winnipeg, Canada R3T 2N2
[3]School of Microelectronics, Xi'an Jiaotong University, Xi'an 710049, People's Republic of China
[4]State Key Laboratory for Mechanical Behavior of Materials, Xi'an Jiaotong University, Xi'an 710049, People's Republic of China

## ABSTRACT

The damping of $La_{0.67}Sr_{0.33}MnO_3$ (LSMO) epitaxial films as a function of thickness at different temperatures was studied. The competition between two scattering types ($\rho$-like and $\sigma$-like) with entirely distinct thickness and temperature dependencies resulted in complicated damping behavior. The behavior of $\sigma$-like damping in LSMO films is consistent with the behavior in magnetic metal films. However, because $\rho$-like damping is sensitive to the fine electron structure near the Fermi surface, the distortion of the oxygen octahedra controlled by the film thickness is an important factor in controlling the damping. Our study demonstrates that the complexity of damping in LSMO epitaxial films is a consequence of strong-correlation effects, which are characteristic of complex transition-metal oxides.

Since the rapid development of spintronics for next-generation memory and processor architectures,[1] the family of spintronic materials has extended from transition metals to complex oxides with strongly correlated electrical and magnetic properties.[2,3] In conventional metals, electron density and spin-orbit coupling are limited owing to their robust electron structure. In contrast, transition-metal oxides with strongly coupled charge, spin, and crystalline structures are promising candidates for improved performance and surprising spintronic effects.[4–6] For example, a two-dimensional electron system with high mobility at oxide interfaces[7] can achieve higher spin-charge interconversion efficiency than conventional heavy metals;[8] current-induced deterministic magnetic field-free switching has been found in all-oxide heterojunctions with higher efficiency.[9]

As a complex transition metal oxide, LSMO film has attracted considerable attention in the field of spintronics because of its high Curie temperature (~360 K) among manganese oxides[10] and a spin polarization rate close to 100%.[11] The latter gives rise not only to an ultra-high tunnel magnetoresistance (1800% at 4 K)[11] but also to an ultra-low Gilbert damping ($5.2 \times 10^{-4}$, grown on $NdGaO_3$ substrate)[12] because of the restricted spin-dependent scattering around the Fermi surface with only one spin-oriented state.[13] In spintronics, Gilbert damping is not only an essential parameter in magnetization dynamics[14] but also limits the threshold current for spin-torque switching and auto oscillators[15,16] and the decay length of diffusive spin waves.[17] However, there is limited research on the damping mechanism of epitaxial LSMO films, and the specifics of their damping tuning mechanism remain controversial. For instance, it is disputed whether the temperature-dependent damping of LSMO films is monotonic.[18,19] To address this discrepancy, our study focused on the influence of strong-correlation effects on the damping mechanism. Specifically, the correlation between the crystalline structure and electronic momentum scattering was systematically studied in epitaxial LSMO films.

LSMO films of different thicknesses ($t$=7.1-38.4 nm) were epitaxially grown by pulsed laser deposition using a KrF laser with a wavelength of 248 nm on (001)-oriented $SrTiO_3$ (STO) substrates. The base pressure was greater than $3.6 \times 10^{-8}$ Torr, and the films were grown at 620 °C with an oxygen pressure of 90 mTorr. The pulse energy was 500 mJ with a frequency of 3 Hz. After deposition, the films were annealed in situ for 1 h at 750 °C under an oxygen atmosphere of 300 Torr, and then cooled to 300 K at a rate of 10 °C/min. The thicknesses of the LSMO films were determined using X-ray reflectometry, and their crystalline structure was characterized using high-resolution X-ray diffraction (XRD) and reflection high-energy electron diffraction (RHEED). Surface topography was characterized using atomic force microscopy (AFM). The magnetic hysteresis loops were measured using a vibrating-sample magnetometer. The temperature-dependent ferromagnetic resonance (FMR) and transport properties were measured using a self-built testing system with a Cryogenic-J4440 cryofree vector magnet.

Figure 1(a) shows the $\theta$-$2\theta$ scan of XRD for LSMO (002) peaks, indicating the LSMO films are epitaxially grown on STO substrates. Local magnifications of the LSMO (002) peaks with thicknesses of 24.0 nm (green) and 38.4 nm (black) are shown in the insets. The clear Laue fringes demonstrate the flatness and uniformity of the epitaxial LSMO film. Moreover, a shift in the LSMO (002) peak position to a higher diffraction angle can be observed, indicating that the values of the out-of-plane lattice constant $c$ decrease with increasing thickness.

To quantify the variation between the lattice constant and thickness, the diffraction angles of LSMO (002) were determined using Jade software, and the values of $c$ were obtained as follows:

$$\frac{2abc\sin\theta}{\sqrt{h^2b^2c^2+k^2a^2c^2+l^2a^2b^2}} = \lambda \quad ,(1)$$

where $a$ and $b$ are the in-plane lattice constants, $h$, $k$, $l$ are Miller indices, and $\lambda$=1.54 Å is the X-ray wavelength. As the diffraction peaks are spread out, the calculated lattice constants are only meaningful as a statistical average. It can be seen from Fig. 1(b) that $c$ decreases with increasing thickness and tends to be saturated. According to A. Vailionis et al.,[20] the variation of the LSMO lattice constant with thickness is due to the spontaneous presence of regions with different lattice constants in the thickness direction of the films. Here, we simplified this complicated situation into a bilayer model: an LSMO interfacial layer with N unit cells and a relatively large $c_i$=3.93 Å, and a bulk-like layer for the remaining part with $c_b$=3.84 Å,[20] where i and b denote interface and bulk-like. Based on this model, the average lattice constant $x$ (=$a$, $b$, $c$) as a function of thickness $t$ is given by

$$\bar{x} = \frac{\mathrm{N}c_i x_i + (t - \mathrm{N}c_i) x_b}{t} \quad . (2)$$

As shown in Fig. 1(b), the fitting curve is consistent with the data and gives $\mathrm{N} \approx 9\mathrm{u.c.}$

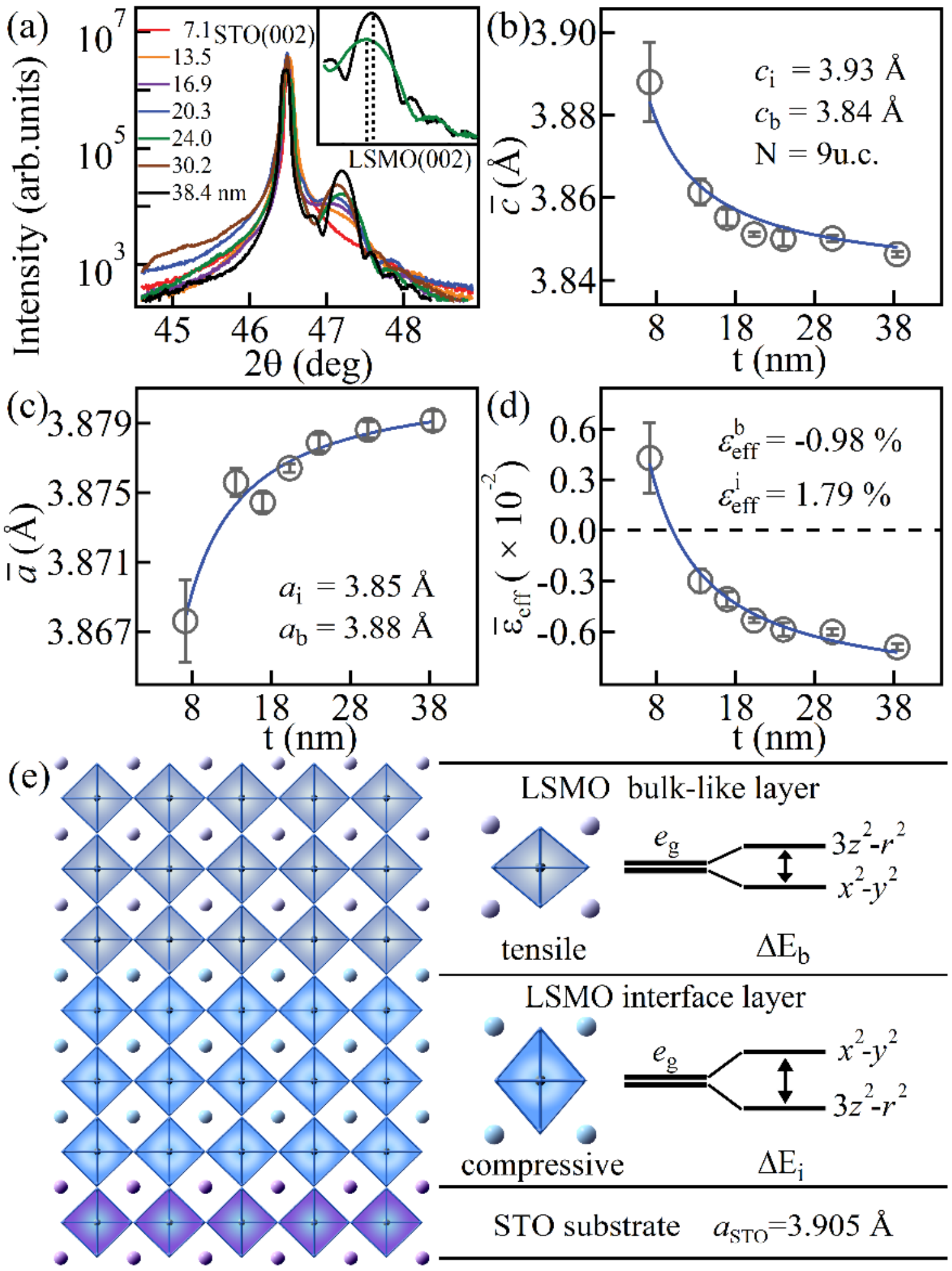

FIG. 1. (a) XRD patterns around the (002) peaks of the films with different thicknesses; the inset is the local magnification of the data for thicknesses of 24.0 nm (green) and 38.4 nm (black). Average (b) out-of-plane, (c) in-plane lattice constants and (d) effective strain as a function of thickness, followed by a fitted curve of Eq. (2). (e) Schematic illustration of the bilayer model, where the oxygen octahedra in the interface layer (bulk-like layer) are compressive (tensile) strained.

To comprehensively evaluate the structural variation of the films with different thicknesses, we also performed XRD analysis of the LSMO (201) peak. Similar to the calculation of $\overline{c}$, the average in-plane lattice constants $\overline{a}$ are shown in Fig. 1(c) as a function of thickness. The parameters $a_{\mathrm{i}}$=3.85 Å, $a_{\mathrm{b}}$=3.88 Å are obtained from the fitting of Eq. (2) with fixing $\mathrm{N} = 9\mathrm{u.c.}$. Compared with the data shown in Fig. 1(b), the values of $\overline{c}$ and $\overline{a}$ are different, exhibiting opposite trends with the thickness. The variation in the LSMO film structure is due to the distortion of the oxygen octahedron, and the degree of distortion varies with thickness. These deformations normally arise from two effects: the Jahn-Teller effect and the strain caused by lattice mismatch.[21,22] As both effects break the symmetry of the oxygen octahedron, we use a total average effective strain $\overline{\varepsilon}_{\mathrm{eff}}$ as the quantitative parameter for describing the structural detail in LSMO films of different thicknesses, given by $\overline{\varepsilon}_{\mathrm{eff}} = \sqrt{\frac{1}{6}}\left(2\overline{\varepsilon}_{zz} - \overline{\varepsilon}_{xx} - \overline{\varepsilon}_{yy}\right)$,[23,24] where $\overline{\varepsilon}_{zz}$ ($\overline{\varepsilon}_{xx}$, $\overline{\varepsilon}_{yy}$) is the out-of-plane (in-plane) strain. The calculations are presented in Fig. 1(d). Based on the bilayer model, the effective strain can be fitted similarly by replacing $x$ with the effective strain in Eq. (2) and the fitting gives $\varepsilon_{\mathrm{eff}}^{\mathrm{i}} = 1.79\%$ ($c>a$, compressive) and $\varepsilon_{\mathrm{eff}}^{\mathrm{b}} = -0.98\%$ ($c<a$, tensile).

As the strong-correlation effect is closely related to the deformation of the oxygen octahedron where $Mn^{3+}$ is located, the details of the crystalline structure should be further investigated. It has been shown that the crystal field caused by the distorted octahedra degenerates the $e_g$ levels of $x^2$-$y^2$ and $3z^2$-$r^2$ orbits.[25,26] In LSMO films, tensile strain favors $x^2$-$y^2$ occupancy, while compressive strain favors $3z^2$-$r^2$ occupancy.[27–29] Consequently, the $3z^2$-$r^2$ ($x^2$-$y^2$) orbit in the interface (bulk-like) layer is preferentially occupied.[29–31] Furthermore, the splitting energy $\Delta E_{\mathrm{i}}$ between the $x^2$-$y^2$ and $3z^2$-$r^2$ orbitals in the interface layer is larger than that of the bulk-like layer $\Delta E_{\mathrm{b}}$, as shown in Fig. 1(e).

Next, we investigated the dynamic magnetic responses of the films using frequency-dependent broadband FMR. Figure 2(a) shows representative FMR spectra obtained with an in-plane [100] magnetic field at room temperature. The spectra were fitted with the universal line-shape equation,[32] allowing us to extract the resonant field $H_0$ and the linewidth $\Delta H,$ as shown in Figs. 2(b) and 2(c), respectively. The vertical shift in Fig. 2(c) was performed to make the data clear. In the absence of significant magnetocrystalline anisotropy in the LSMO films studied here [Supplemental Material IV, Figs. S4(a)-(b)], the relation between $H_0$ and the microwave angular frequency $\omega$ is given by $\omega = \gamma\sqrt{(H_0 + 4\pi M_{\mathrm{eff}})H_0}$, where $\gamma/2\pi$ =2.8 GHz/kOe is the gyromagnetic ratio.[33] We determined the effective Gilbert damping $\alpha_{\mathrm{eff}}$ by fitting $\Delta H$ as a function of $\omega$, which is given by:

$$\Delta H = \Delta H_0 + \frac{\alpha_{\mathrm{eff}}\omega}{\gamma} \quad .(3)$$

Here, $\Delta H_0$ represents the extrinsic linewidth, which can originate from defects and magnetic inhomogeneities. The effective Gilbert damping $\alpha_{\mathrm{eff}}$ consists of intrinsic and extrinsic damping.[34,35] In single-layer films, the latter is typically attributed to two-magnon scattering (TMS), eddy current, radiative, and magnetoelastic damping. We estimated the maximum contributions of eddy-current, radiative, and magnetoelastic damping to be ≈1.1% (30.2 nm at 50 K), 1.8% (38.4 nm at 50 K), and 3.0% (7.1 nm at 50 K) of $\alpha_{\mathrm{eff}}$, respectively (Supplemental Material II). Therefore, we can reasonably neglect these contributions in the following discussion.

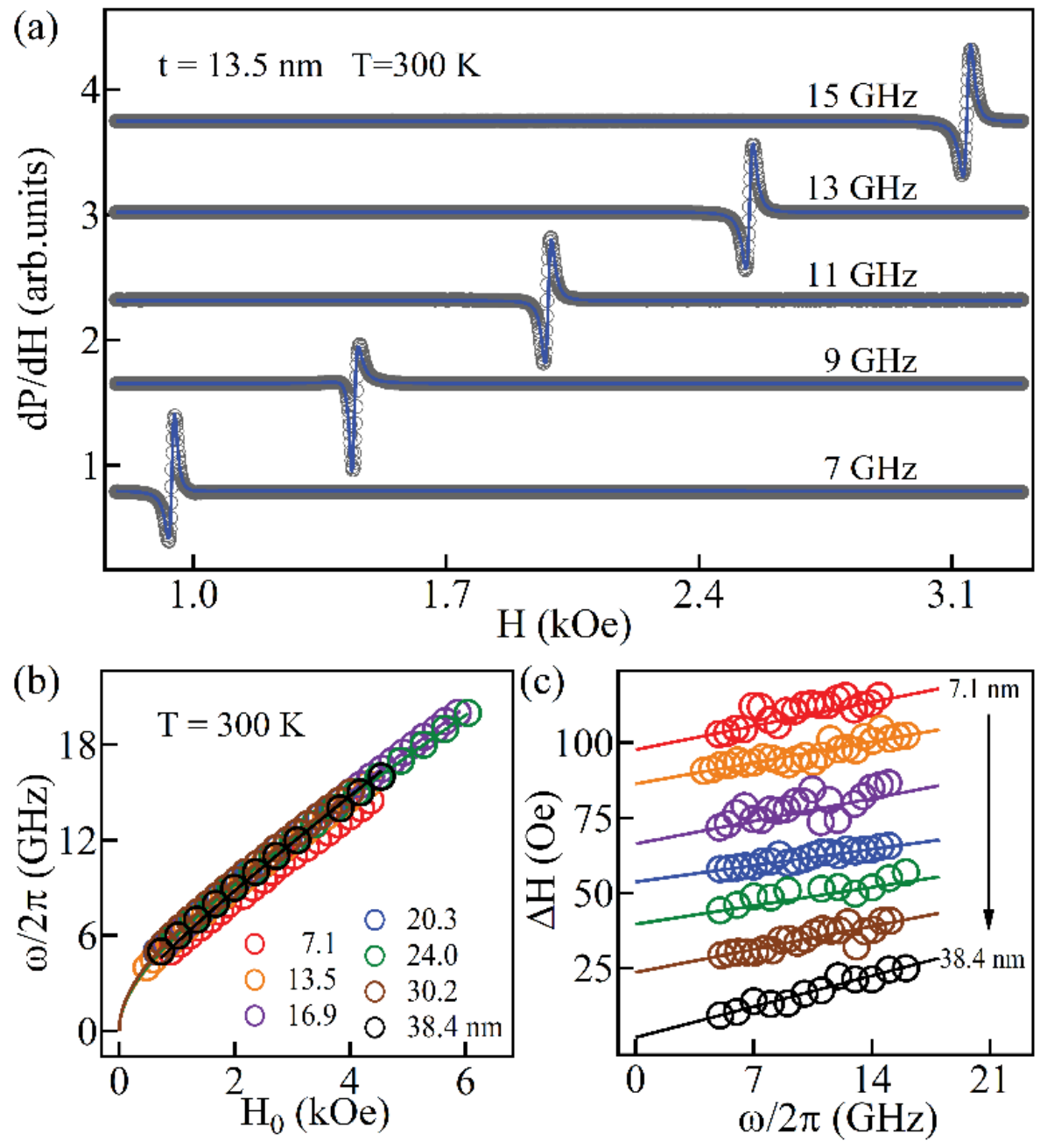

FIG. 2. (a) Typical FMR spectra with applied in-plane magnetic field. (b) Microwave angular frequency versus resonant field; solid lines are fitting of Kittel's formula. (c) Frequency-dependent resonance linewidth; the vertical shift was performed to make the data clear.

According to Xu et al., the disentanglement of various damping mechanisms in ferromagnetic metal films can be achieved by measuring the thickness-dependent $\alpha_{\mathrm{eff}}$, which is given by

$$\alpha_{\mathrm{eff}} = \alpha_b + \frac{\beta_i}{t} + \frac{\beta_{\mathrm{TMS}}}{t^2} \tag{4}$$

,[35] where $\alpha_{\mathrm{b}}$ is the bulk damping, $\beta_{\mathrm{i}}$ and $\beta_{\mathrm{TMS}}$ are the coefficients of interface damping (including spin pumping for multilayers) and TMS, respectively. Figure 3(a) shows the variation of $\alpha_{\mathrm{eff}}$ with thickness at room temperature, which exhibits a non-monotonic trend and is inconsistent with Eq. (4). As the temperature decreases, the thickness-dependent damping gradually becomes monotonic [Figs. 3(a)–(c)], and the model in Eq. (4) is adequate for fitting the data when the temperature is below 100 K [Supplemental Material III, Fig. S2(e)].

According to previous studies on the damping of magnetic metal films,[35–37] the thickness-dependent damping is expected to be consistent with Eq. (4) at different temperatures. Consequently, the data in Figs. 3(a)–(c) suggest a complicated damping mechanism for LSMO films. Building upon prior discussions, the crystalline structure of the LSMO film exhibited variations in thickness, as indicated in Fig. 1(d). Despite the relatively small deformation of the crystalline structure, it has a significant influence on the electronic structures of strongly correlated systems.[4] As a result, significant changes in the spin-related scattering in the LSMO film induced by spin-orbit coupling are anticipated. Therefore, a comprehensive understanding of the coupling between the structure and damping necessitates the separation of the damping contributed by different types of scattering.

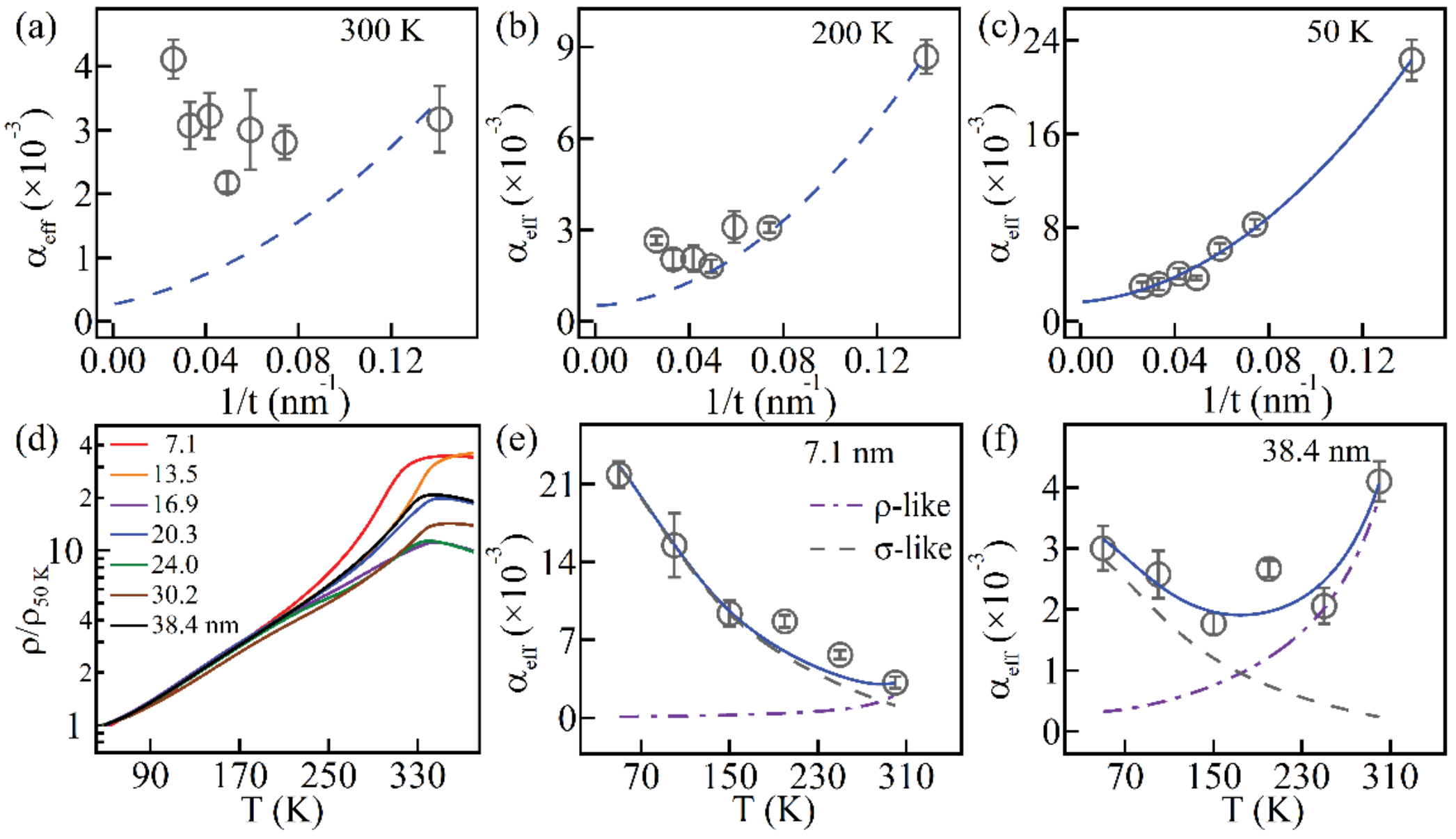

FIG. 3. Thickness-dependent $\alpha_{\mathrm{eff}}$ of LSMO films at (a) 300 K, (b) 200 K and (c) 50K. Blue dashed lines and solid lines represent the guided and fitted curves of Eq. (4), respectively. (d) Temperature-dependent normalized resistivity of LSMO films. All of the samples possess ferromagnetic order in this work. Symbols are temperature-dependent damping in (e) 7.1 nm and (f) 38.4 nm LSMO films, solid lines are fitting curves of Eq. (5), purple dotted lines and gray dashed lines represent the contribution of $\rho$-like and $\sigma$-like damping.

Kambersky's torque correlation model[38,39] provides a framework for partitioning the Gilbert damping in ferromagnetic transition metals into two scattering types: "conductivity-like" ($\sigma$-like) damping and "resistivity-like" ($\rho$-like) damping, which arise from intraband and interband scattering, respectively. The thickness-dependent effective damping parameter $\alpha_{\mathrm{eff}}$ governed by Eq. (4) becomes increasingly significant as the temperature decreases [Figs. 3(a)–(c)], similar to the behavior of $\sigma$-like damping. Therefore, by referring to this method, we phenomenologically separated the temperature-dependent $\alpha_{\mathrm{eff}}$ using different types of scattering:

$$\alpha_{\mathrm{eff}} = \alpha_\rho(t)\frac{\rho(T)}{\rho(300\ \mathrm{K})} + \alpha_\sigma(t)\frac{\sigma(T)}{\sigma(300\ \mathrm{K})} \quad ,(5)$$

[17,18,35] where $\alpha_\rho(t)$ and $\alpha_\sigma(t)$ represent the interband and intraband scattering parameters, and $\rho(300\ \mathrm{K})$ and $\sigma(300\ \mathrm{K})$ are the resistivity and conductivity at 300 K.

Figure 3(d) displays the temperature-dependent normalized resistivity, which indicates the promising half-metallic nature of LSMO films. Using the temperature-dependent resistivity, the effective damping is separated into two distinct contributions by Eq. (5). The fitting result is illustrated in Figs. 3(e)-(f). The results show that the $\rho$-like ($\sigma$- like) damping dominates $\alpha_{\text{eff}}$ at high (low) temperature [Figs. 3(e)-(f)].

The $\sigma$-like damping $[\alpha_\sigma(t,T) \equiv \alpha_\sigma(t)\frac{\sigma(T)}{\sigma(300\text{ K})}]$ is shown in Figs. 4(a)–(c) as a function of $1/t$ at different temperatures. $\alpha_\sigma(t,T)$ is linear with $1/t$ at 300 K [Fig. 4(a)], in line with prior reports.[12] The degree of nonlinearity in $\alpha_\sigma(t,T)$ becomes more evident with decreasing temperature [Figs. 4(b) and (c)]. This suggests that the TMS becomes more significant with decreasing temperature. In contrast to $\alpha_{\text{eff}}$, the thickness-dependent $\alpha_\sigma(t,T)$ shows agreement with Eq. (4) at different temperatures [blue line in Figs. 4(a)–(c)], which is consistent with previous studies on ferromagnetic metal films.[35–37] The contributions of bulk $\alpha_{\text{b}}$, interface $\beta_{\text{i}}$, and TMS $\beta_{\text{TMS}}$ extracted from $\alpha_\sigma(T,t)$ is shown in Figs. 4(d)-(f).

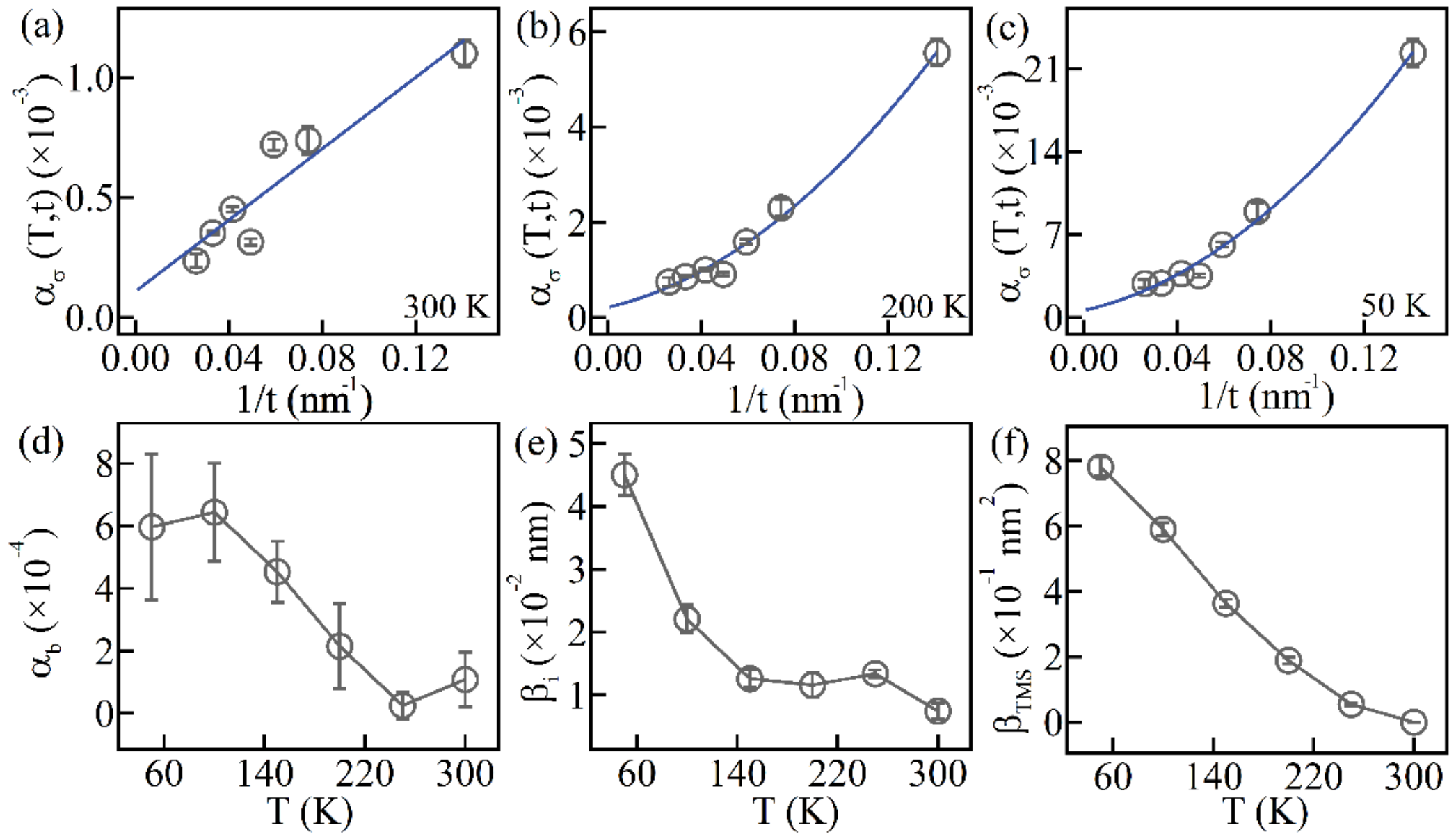

FIG. 4. Thickness-dependent $\sigma$-like damping at (a) 300 K, (b) 200 K and (c) 50K. Blue lines are fitted curves of Eq. (4). Temperature-dependent (d) bulk $\alpha_{\text{b}}$, (e) interface $\beta_{\text{i}}$ and (f) TMS $\beta_{\text{TMS}}$ contribution to $\sigma$-like damping.

As shown in Fig. 4(d), $\alpha_{\text{b}}$ is quite small ($10^{-4}$); hence, the enhancement of $\alpha_\sigma(T,t)$ at low temperatures can be attributed to the contributions of the interface (28% for 7.1 nm at 50 K) and TMS (69%). The temperature-dependent $\beta_{\text{i}}$ shown in Fig. 4(e) increases with decreasing temperature. According to Haspot et al. ,[18] in LSMO ultrathin films, the precession of magnetization injects a pure spin current into the magnetically active dead layer and contributes to damping, which increases with decreasing temperature. Thus, we attribute the interface contribution $\beta_{\text{i}}$ to spin pumping by the magnetically active dead layer. The linear proportionality of the temperature-dependent $\alpha_\sigma(t,T)$ with $1/t$ at 300 K [Fig. 4(a)] and the trend of $\beta_{\mathrm{TMS}}$ with decreasing temperature [Fig. 4(f)] indicate the transition from negligible to significant TMS contribution as the temperature decreases. To confirm the negligible TMS, we conducted in-plane angle-dependent electron spin resonance with 9.46 GHz at 300 K and estimated the TMS contribution accounts for $\approx$3.7% of $\alpha_\sigma$(300 K, 9.7 nm) [Supplemental Material IV, Fig. S4(e)]. Furthermore, we attributed the trend of $\beta_{\mathrm{TMS}}$ increasing with decreasing temperature to the enhanced anisotropic field at the interface, as reported by Xu et al. .[35]

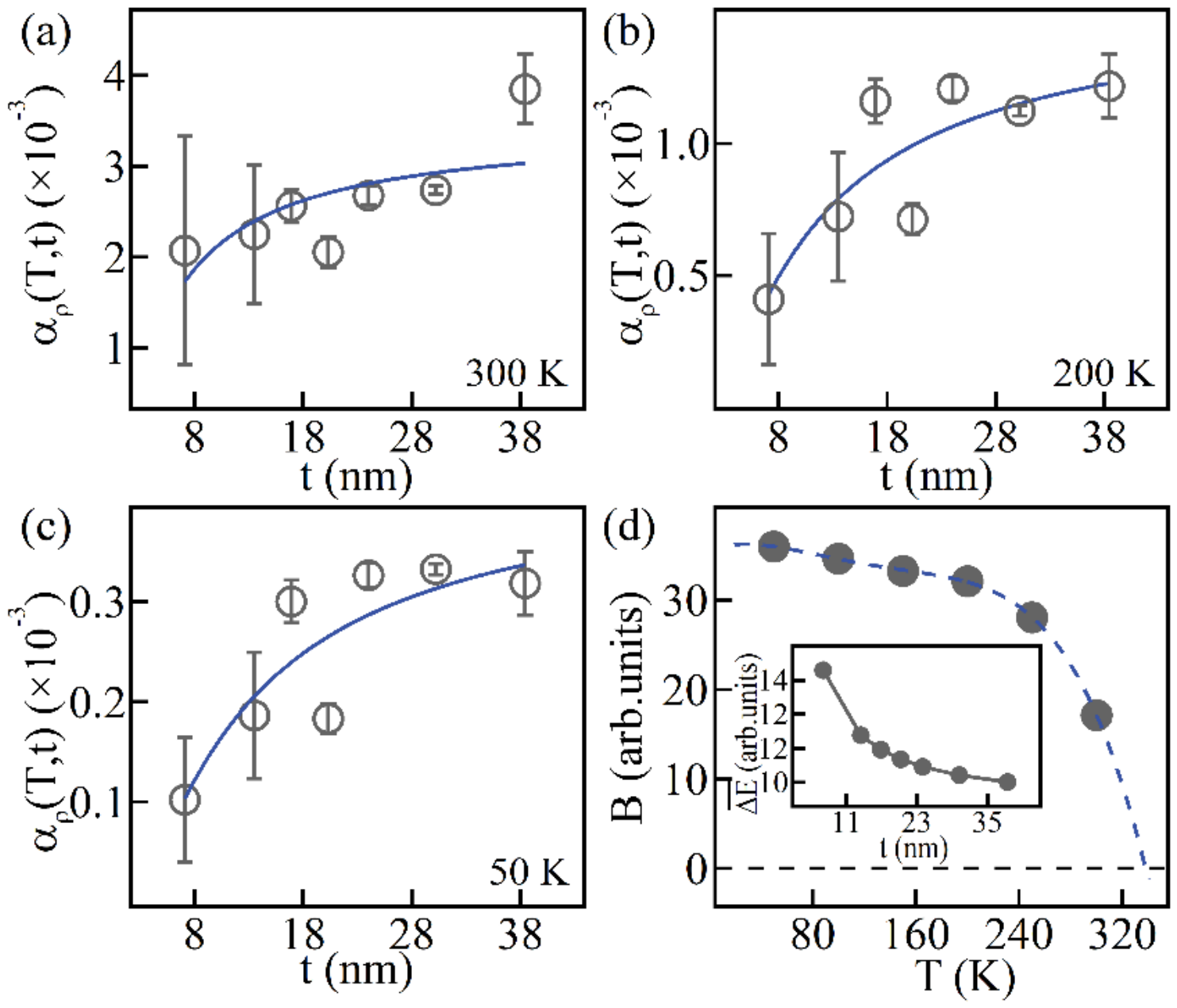

FIG. 5. Thickness-dependent $\rho$-like damping (a) 300 K, (b) 200 K and (c) 50K. Blue lines are fitted curves of Eq. (6). (d) The fitting parameter $B(T)$ depending on temperature. The inset shows a thickness-dependent average energy gap of $e_g$ level splitting calculated from $\overline{\Delta E}(t) \propto \frac{Nc_i|\varepsilon_{\mathrm{eff}}^i|+(t-Nc_i)|\varepsilon_{\mathrm{eff}}^b|}{t}$ based on the bilayer model in LSMO films.

We next present a detailed discussion on thickness-dependent $\rho$-like damping [ $\alpha_\rho(t,T) \equiv \alpha_\rho(t)\frac{\rho(T)}{\rho(300\ \mathrm{K})}$ ]. In Figs. 5(a)-(c), $\alpha_\rho(t,T)$ increases with increasing thickness, unlike $\alpha_\sigma(t,T)$. For interband scattering, the magnetization dynamics excite electron-hole pairs across different bands, which is related to the fine electronic structure near the Fermi surface.[40,41] Therefore, the increase in $\alpha_\rho(t,T)$ with thickness may be attributed to the differences in the fine electronic structure near the Fermi surface resulting from thickness variation. Thus, we speculate that the damping resulting from interband scattering is influenced by the energy gap of the $e_g$ level splitting due to octahedral deformation. Within the bilayer model, the average energy gap of the split orbital is proportional to the effective strain magnitude, i.e., $\overline{\Delta E}(t) \propto |\bar{\varepsilon}_{\mathrm{eff}}| = \frac{Nc_{\mathrm{i}}|\varepsilon_{\mathrm{eff}}^i|+(t-Nc_{\mathrm{i}})|\varepsilon_{\mathrm{eff}}^b|}{t}$ [see insert of Fig. 5(d)].Therefore, the phenomenological

expression for $\alpha_\rho(t,T)$ as a function of thickness can be given by

$$\alpha_\rho(t,T) = \mathrm{A}e^{-\overline{\Delta E}(t)/B(T)} \tag{6}$$

where A is a normalized parameter and $B(T)$ characterizes the energy required to excite electron interband transitions at different temperatures. The results in Figs. 5(a)-(c) show that this simple deduction fits well with the experimental data, indicating that $\alpha_\rho(t,T)$ is related to the deformation of the oxygen octahedra. This is a manifestation of the strong-correlation effect in tuning damping in LSMO films. Additionally, the $B(T)$ shown in Fig. 5(d), which saturates at low temperatures, decreases with increasing temperature, and approaches zero near the Curie temperature of LSMO (7.1 nm, ~340K). The dependence of $B(T)$ on temperature is similar to that of magnetization. Note that, the floor level energy of spin waves is proportional to $\hbar\omega = \hbar\gamma\sqrt{H_0(H_0 + 4\pi M_{\mathrm{eff}})}$, which is directly linked to the effective demagnetization field $4\pi M_{\mathrm{s}}$. Hence, the energy required to excite electron interband transitions may be related to magnons through spin-orbit coupling.

Therefore, the non-monotonic trend of $\alpha_{\mathrm{eff}}$ at high temperatures [Figs. 3(a)-(b)] is caused by the pronounced $\alpha_\rho(t,T)$. Furthermore, both the interface and TMS contributions explain why the thickness-dependent $\alpha_{\mathrm{eff}}$ shown in Figs. 3(a)–(c) gradually agree with Eq. (4), because $\alpha_\sigma(t,T)$ is dominant at low temperature. Based on the suppression of the contribution of $\alpha_\rho(t,T)$ to $\alpha_{\mathrm{eff}}$ with decreasing temperature, we conclude that the different trends of thickness-dependent $\alpha_{\mathrm{eff}}$ at different temperatures is the result of competition between $\alpha_\sigma(t,T)$ and $\alpha_\rho(t,T)$, which have entirely distinct thickness and temperature dependences.

In summary, we studied the damping mechanism of LSMO epitaxial films by thickness-dependent $\rho$-like and $\sigma$-like damping at different temperatures. $\alpha_\sigma(t,T)$ consists of bulk, interface, and TMS contributions, which decrease with increasing thickness in line with the literature. However, owing to the strong-correlation effects, variations in the crystalline structure of LSMO can significantly influence the fine electronic structure near the Fermi surface to tune the $\alpha_\rho(t,T)$, which combined with the competition with $\alpha_\sigma(t,T)$ results in the complexity of the LSMO films damping. Through spin-orbit coupling, the energy required to excite electron interband transitions may arise from the magnon induced by magnetization dynamics. Our work demonstrates that in spintronics it is necessary to understand the coupling between the crystalline and electronic structures of complex transition metal oxides.

See Supplemental Material that includes information on Surface topography and static magnetic properties of LSMO films (Sect. I), the calculated detail of eddy-current, radiative and magnetoelastic damping for eddy-current (Sect. II), the discussion of the temperature-dependence of $\alpha_{\mathrm{eff}}$ (Sect. III) and in-plane angle-dependent electeon spin resonance (Sect. IV).

This project was supported by National Natural Science Foundation of China (Grant No. 52171231), the Program for Changjiang Scholars and Innovative Research Team in University (Grant No. IRT-16R35), and the 111 Project under Grant No. B20063. We would like to thank Editage (www.editage.cn) for English language editing.

## AUTHOR DECLARATIONS

### Conflict of Interest

The authors have no conflicts to disclose.

### Author Contributions

**Yifei Wang**: Data curation (lead); Investigation (lead); Methodology (lead); Writing - original draft (lead). **Xinxin Fan**: Methodology (equal); Writing – review & editing (supporting). **Xiaoyu Feng**: Investigation (supporting); Writing – review & editing (supporting). **Xiaohu Gao**: Writing – review & editing (supporting). **Yunfei Ke**: Writing – review & editing (supporting). **Jiguang Yao**: Investigation (supporting). **Muhan Guo**: Writing – review & editing (supporting). **Tao Wang**: Supervision (equal); Writing – review & editing (supporting). **Lvkang Shen**: Supervision (equal). **Ming Liu**: Supervision (equal). **Desheng Xue**: Supervision (equal). **Xiaolong Fan**: Project administration (lead); Supervision (lead); Writing – review & editing (lead).

## DATA AVAILABILITY

The data that support the findings of this study are available from the corresponding author upon reasonable request.

## REFERENCES

[1] J. Varignon, L. Vila, A. Barthélémy, and M. Bibes, Nat. Phys. **14**(4), 322–325 (2018).

[2] M. Bibes, and A. Barthelemy, IEEE Trans. Electron Devices **54**(5), 1003–1023 (2007).

[3] S. Majumdar, and S. van Dijken, J. Phys. D. Appl. Phys. **47**(3), 34010 (2014).

[4] H.Y. Hwang, Y. Iwasa, M. Kawasaki, B. Keimer, N. Nagaosa, and Y. Tokura, Nat. Mater. **11**(2), 103–113 (2012).

[5] M. Imada, A. Fujimori, and Y. Tokura, Rev. Mod. Phys. **70**(4), 1039–1263 (1998).

[6] P. Zubko, S. Gariglio, M. Gabay, P. Ghosez, and J.-M. Triscone, Annu. Rev. Condens. Matter Phys. **2**(1), 141–165 (2011).

[7] A. Ohtomo, and H.Y. Hwang, Nature **427**(6973), 423–426 (2004).

[8] E. Lesne, Y. Fu, S. Oyarzun, J.C. Rojas-Sánchez, D.C. Vaz, H. Naganuma, G. Sicoli, J.-P. Attané, M. Jamet, E. Jacquet, J.-M. George, A. Barthélémy, H. Jaffrès, A. Fert, M. Bibes, and L. Vila, Nat. Mater. **15**(12), 1261–1266 (2016).

[9] L. Liu, Q. Qin, W. Lin, C. Li, Q. Xie, S. He, X. Shu, C. Zhou, Z. Lim, J. Yu, W. Lu, M. Li, X. Yan, S.J. Pennycook, and J. Chen, Nat. Nanotechnol. **14**(10), 939–944 (2019).

[10] A. Urushibara, Y. Moritomo, T. Arima, A. Asamitsu, G. Kido, and Y. Tokura, Phys. Rev. B **51**(20), 14103–

14109 (1995).

[11] M. Bowen, M. Bibes, A. Barthélémy, J.-P. Contour, A. Anane, Y. Lemaître, and A. Fert, Appl. Phys. Lett. **82**(2), 233–235 (2003).

[12] Q. Qin, S. He, W. Song, P. Yang, Q. Wu, Y.P. Feng, and J. Chen, Appl. Phys. Lett. **110**(11), 112401 (2017).

[13] J.-H. Park, E. Vescovo, H.-J. Kim, C. Kwon, R. Ramesh, and T. Venkatesan, Nature **392**(6678), 794–796 (1998).

[14] B. Heinrich, *Ultrathin Magnetic Structures III* (Springer-Verlag, Berlin, Heidelberg, 2005).

[15] D.C. Ralph, and M.D. Stiles, J. Magn. Magn. Mater. **320**(7), 1190–1216 (2008).

[16] A. Brataas, A.D. Kent, and H. Ohno, Nat. Mater. **11**(5), 372–381 (2012).

[17] B. Khodadadi, A. Rai, A. Sapkota, A. Srivastava, B. Nepal, Y. Lim, D.A. Smith, C. Mewes, S. Budhathoki, A.J. Hauser, M. Gao, J.-F. Li, D.D. Viehland, Z. Jiang, J.J. Heremans, P. V Balachandran, T. Mewes, and S. Emori, Phys. Rev. Lett. **124**(15), 157201 (2020).

[18] V. Haspot, P. Noël, J.-P. Attané, L. Vila, M. Bibes, A. Anane, and A. Barthélémy, Phys. Rev. Mater. **6**(2), 024406 (2022).

[19] V. Flovik, F. Macià, S. Lendínez, J.M. Hernàndez, I. Hallsteinsen, T. Tybell, and E. Wahlström, J. Magn. Magn. Mater. **420**, 280–284 (2016).

[20] A. Vailionis, H. Boschker, Z. Liao, J.R.A. Smit, G. Rijnders, M. Huijben, and G. Koster, Appl. Phys. Lett. **105**(13), 131906 (2014).

[21] E. Dagotto, *Nanoscale Phase Separation and Colossal Magnetoresistance* (Springer, Berlin, 2003).

[22] O. Diéguez, K.M. Rabe, and D. Vanderbilt, Phys. Rev. B **72**(14), 144101 (2005).

[23] P. Dey, T.K. Nath, and A. Taraphder, Appl. Phys. Lett. **91**(1), 12511 (2007).

[24] F. Tsui, M.C. Smoak, T.K. Nath, and C.B. Eom, Appl. Phys. Lett. **76**(17), 2421–2423 (2000).

[25] Z. Fang, I. V Solovyev, and K. Terakura, Phys. Rev. Lett. **84**(14), 3169–3172 (2000).

[26] H. Zenia, G.A. Gehring, G. Banach, and W.M. Temmerman, Phys. Rev. B **71**(2), 24416 (2005).

[27] C. Aruta, G. Ghiringhelli, A. Tebano, N.G. Boggio, N.B. Brookes, P.G. Medaglia, and G. Balestrino, Phys. Rev. B **73**(23), 235121 (2006).

[28] M. Huijben, L.W. Martin, Y.-H. Chu, M.B. Holcomb, P. Yu, G. Rijnders, D.H.A. Blank, and R. Ramesh, Phys. Rev. B **78**(9), 94413 (2008).

[29] A. Tebano, A. Orsini, P.G. Medaglia, D. Di Castro, G. Balestrino, B. Freelon, A. Bostwick, Y.J. Chang, G. Gaines, E. Rotenberg, and N.L. Saini, Phys. Rev. B **82**(21), 214407 (2010).

[30] D. Pesquera, G. Herranz, A. Barla, E. Pellegrin, F. Bondino, E. Magnano, F. Sánchez, and J. Fontcuberta, Nat. Commun. **3**(1), 1189 (2012).

[31] A. Tebano, C. Aruta, S. Sanna, P.G. Medaglia, G. Balestrino, A.A. Sidorenko, R. De Renzi, G. Ghiringhelli, L. Braicovich, V. Bisogni, and N.B. Brookes, Phys. Rev. Lett. **100**(13), 137401 (2008).

[32] E. Montoya, T. McKinnon, A. Zamani, E. Girt, and B. Heinrich, J. Magn. Magn. Mater. **356**, 12–20 (2014).

[33] C. Kittel, Phys. Rev. **73**(2), 155–161 (1948).

[34] P. He, X. Ma, J.W. Zhang, H.B. Zhao, G. Lüpke, Z. Shi, and S.M. Zhou, Phys. Rev. Lett. **110**(7), 077203 (2013).

[35] Z. Xu, K. Zhang, and J. Li, Phys. Rev. B **104**(22), 224404 (2021).

[36] Y. Zhao, Q. Song, S.-H. Yang, T. Su, W. Yuan, S.S.P. Parkin, J. Shi, and W. Han, Sci. Rep. **6**(1), 22890 (2016).

[37] G. Lu, X. Huang, S. Fan, W. Ling, M. Liu, J. Li, L. Jin, and L. Pan, J. Alloys Compd. **753**, 475–482 (2018).

[38] V. Kamberský, Czechoslov. J. Phys. B **26**(12), 1366–1383 (1976).

[39] V. Kamberský, Can. J. Phys. **48**(24), 2906–2911 (1970).

[40] K. Gilmore, Y.U. Idzerda, and M.D. Stiles, Phys. Rev. Lett. **99**(2), 27204 (2007).

[41] X. Ma, L. Ma, P. He, H.B. Zhao, S.M. Zhou, and G. Lüpke, Phys. Rev. B **91**(1), 014438 (2015).